\documentclass[journal=jcisd8,manuscript=article,layout=twocolumn,email=true]{achemso}

\usepackage[version=3]{mhchem} % Formula subscripts using \ce{}
\usepackage[T1]{fontenc}       % Use modern font encodings
\usepackage{amsmath,amssymb,exscale}
\usepackage{multirow}
\usepackage{booktabs}
\usepackage{threeparttable}
\usepackage{algorithm}
\usepackage{algpseudocode,amsmath}
\usepackage{pifont}

\author{Janaina Cruz Pereira}
\affiliation[Oswaldo Cruz Foundation]
{Fiocruz, 4365 Avenida Brasil, Rio de Janeiro, RJ 21040 900, Brazil}
\email {janaina.pereira@ioc.fiocruz.br}
\author{Ernesto Ra\'{u}l Caffarena}
\affiliation[Oswaldo Cruz Foundation]
{Fiocruz, 4365 Avenida Brasil, Rio de Janeiro, RJ 21040 900, Brazil}
\email{ernesto@fiocruz.br}
\author{Cicero Nogueira dos Santos}
\email{cicerons@us.ibm.com}
\affiliation[IBM Watson]
{IBM Watson, 1101 Kitchawan Rd, Yorktown Heights, NY 10598, USA}

\title[Deep learning-based method for virtual screening]
  {Boosting Docking-based Virtual Screening with Deep Learning}

\keywords{Drug discovery, Deep Convolutional Neural Networks and Docking Data}

\begin{document}

% \begin{tocentry}

% Some journals require a graphical entry for the Table of Contents.
% This should be laid out ``print ready'' so that the sizing of the

% Inside the \texttt{tocentry} environment, the font used is Helvetica
% 8\,pt, as required by \emph{Journal of the American Chemical

% The surrounding frame is 9\,cm by 3.5\,cm, which is the maximum
% permitted for  \emph{Journal of the American Chemical Society}
% graphical table of content entries. The box will not resize if the
% content is too big: instead it will overflow the edge of the box.

% This box and the associated title will always be printed on a
% separate page at the end of the document.

%	\includegraphics  {TOC_graphic_9cm_35cm.eps}

 %\end{tocentry}

\begin{abstract}

In this work,
we propose a deep learning approach to improve docking-based virtual screening.
The introduced deep neural network,
DeepVS,
uses the output of a docking program and learns how to extract relevant features from basic data such as atom and residues types
obtained from protein-ligand complexes.
Our approach introduces the use of atom and amino acid embeddings and implements an effective way of creating distributed vector representations of protein-ligand complexes by modeling the compound as a set of atom contexts that is further processed by a convolutional layer.
One of the main advantages of the proposed method is that it does not require feature engineering.
We evaluate DeepVS on the Directory of Useful Decoys (DUD),
using the output of two docking programs:
AutodockVina1.1.2 and Dock6.6.
Using a strict evaluation with leave-one-out cross-validation,
DeepVS outperforms the docking programs in both AUC ROC and enrichment factor.
Moreover,
using the output of AutodockVina1.1.2,
DeepVS achieves an AUC ROC of 0.81,
which,
to the best of our knowledge, 
is the best AUC reported so far for virtual screening using the 40 receptors from DUD.

\end{abstract}

\section{Introduction}
Drug discovery process is a time-consuming and expensive task. The development and even the repositioning of already known compounds is a difficult chore \cite{Hecht2009}. The scenario gets worse if we regard the thousands or millions of molecules capable of being synthesized in each development stage \cite{Walters1998,Cheng2012}.

In the past, experimental methods such as high-throughput screening (HTS) could help making this decision through the screening of large chemical libraries against a biological target. However, the high cost of the whole process associated with a low success rate turns this method inaccessible to the academia \cite{Walters1998,Cheng2012,Shoichet2004}.

In order to overcome these difficulties, the use of low-cost computational alternatives is extensively encouraged, and it was adopted routinely as a way to aid in the development of new drugs \cite{Cheng2012,Shoichet2004,Ghosh2006}.

Computational virtual screening works basically as a filter (or a prefilter) consisting of the virtual selection of molecules, based on a particular predefined criterion of potentially active compounds against a determined pharmacological target \cite{Walters1998,Shoichet2004,Bissantz2000}.

Two variants of this method can be adopted: ligand-based virtual screening and structure-based virtual screening. The first one deals with the similarity and the physicochemical analysis of active ligands to predict the activity of other compounds with similar characteristics.  The second is utilized when the three-dimensional structure of the target receptor was already elucidated somehow (experimentally or computationally modeled). This approach is used to explore molecular interactions between possible active ligands and residues of the binding site. Structure-based methods present a better performance when compared to methods based solely on the structure of the ligand focusing on the identification of new compounds with therapeutic potential \cite{Arciniega2014,Drwal2013,Hecht2009,Schneider2010}.

One of the computational methodologies extensively used to investigate these interactions is molecular docking \cite{Ghosh2006,Drwal2013,Kitchen2004}. The selection of more potent ligands using Docking-based Virtual Screening (DBVS) is made by performing the insertion of each compound from a compound library into a particular region of a target receptor with elucidated 3D structure. In the first stage of this process, a heuristic search is carried out in which thousands of possibilities of insertions are regarded. In the second, the quality of the insertion is described via some mathematical functions (scoring functions) that give a clue about the energy complementarity between compound and target \cite{Kitchen2004,Durrant2013}. The last phase became a challenge to the computational scientists considering that it is easier to recover the proper binding mode of a compound within an active site than to assess a low energy score to a determined pose. This hurdle constitutes a central problem to the docking methodology \cite{Kinnings2011}.

Systems based on machine learning (ML) have been successfully used to improve the outcome of Docking-based Virtual Screening for both, increasing the performance of score functions and constructing binding affinity classifiers \cite{Cheng2012,Hecht2009}. The main strategies used in virtual screening are neural networks - NN \cite{Durrant2010}, support vector machines - SVM \cite{Kinnings2011} and random forest - RF \cite{Ballester2010}. 
One of the main advantages of employing ML is the capacity to explain the non-linear dependence of the molecular interactions between ligand and receptor \cite{Cheng2012}.

Traditional Machine Learning strategies depend on how the data is presented. For example, in the virtual screening approaches, scientists normally analyze the docking output to generate or extract human engineered features. Although this process can be effective to some degree, the manual identification of characteristics is a laborious and complex process and can not be applied in large scale, resulting in the loss of relevant information, consequently leading to a set of features incapable of explaining the actual complexity of the problem \cite{Bengio2013,Lecun2015,Cheng2012,Hecht2009}.

On the other hand,
recent work on Deep Learning (DL),
a family of ML approaches that minimizes feature engineering,
has demonstrated enormous success in different tasks from multiple fields\cite{Bengio2013,Lecun2015,Bengio2009}.
DL approaches normally learn features (representations) directly from the raw data with minimal or none human intervention,
which makes the resulting system easier to adapt to new datasets.

In the last few years, 
DL is bringing the attention of the academic community and big pharmaceutical industries, as a viable alternative to aid in the discovery of new drugs. One of the first work in which DL was successfully applied solved problems related to QSAR (Quantitative Structure-Activity Relationships) by Merck in 2012. Some years later, Dahl \latin{et al.} \cite{Dahl2014} developed a multi-task deep neural network to predict biological and chemical properties of a compound directly from its molecular structure. More recently, Multi-task deep neural networks were employed to foresee the active-site directed pharmacophore and toxicity \cite{Unterthiner2014,Unterthiner2015}.  Also in 2015, Ramsundar \latin{et al.} \cite{Ramsundar2015} predicted drug activity using Massively Multitask Neural Networks associated with fingerprints.
The relevance of DL is also highlighted by
recent applications to the discovery of new drugs and the determination of their characteristics such as Aqueous Solubility prediction \cite{Lusci2013} and  Fingerprints \cite{Duvenaud2015}.

In this work,
we propose an approach based on Deep Convolutional Neural Networks to improve docking-based virtual screening.
The method uses docking simulation results as input to a Deep Neural Network, 
DeepVS from now on,
which automatically learns to extract relevant features from basic data such as 
compound atom types,
atomic partial charges,
and the distance between atoms.
DeepVS learns abstract features that are suitable to discriminate between active ligands and decoys in a protein-compound complex.
To the best of our knowledge,
this work is the first on using deep learning to improve docking-based virtual screening.
Recent works on improving docking-based virtual screening have only used traditional shallow neural networks with human defined features \cite{Durrant2011,Durrant2010,Arciniega2014}.

We evaluated DeepVS on the Directory of Useful Decoys,
which contains 40 different receptors.
In our experiments we used the output of two docking programs:
AutodockVina1.1.2 and Dock6.6.
DeepVS outperformed the docking programs in both AUC ROC and enrichment factor.
Moreover,
when compared to the results of other systems previously reported in the literature,
DeepVS achieved the state-of-the-art AUC of 0.81.

The main contributions of this work are:
(1) Proposition of a new deep learning-based approach that achieves state-of-the-art performance on docking-based virtual screening;
(2) Introduction of the concept of atom and amino acid embeddings,  which can also be used in other deep learning applications for computational biology;
(3) Proposition of an effective method to create distributed vector representations of protein-compound complexes that models the compound as a set of atom contexts that is further processed by a convolutional layer.

\section{Materials and Methods}

\subsection{DeepVS}

\begin{figure}[!htb]
	\centering
	\includegraphics[scale=1]{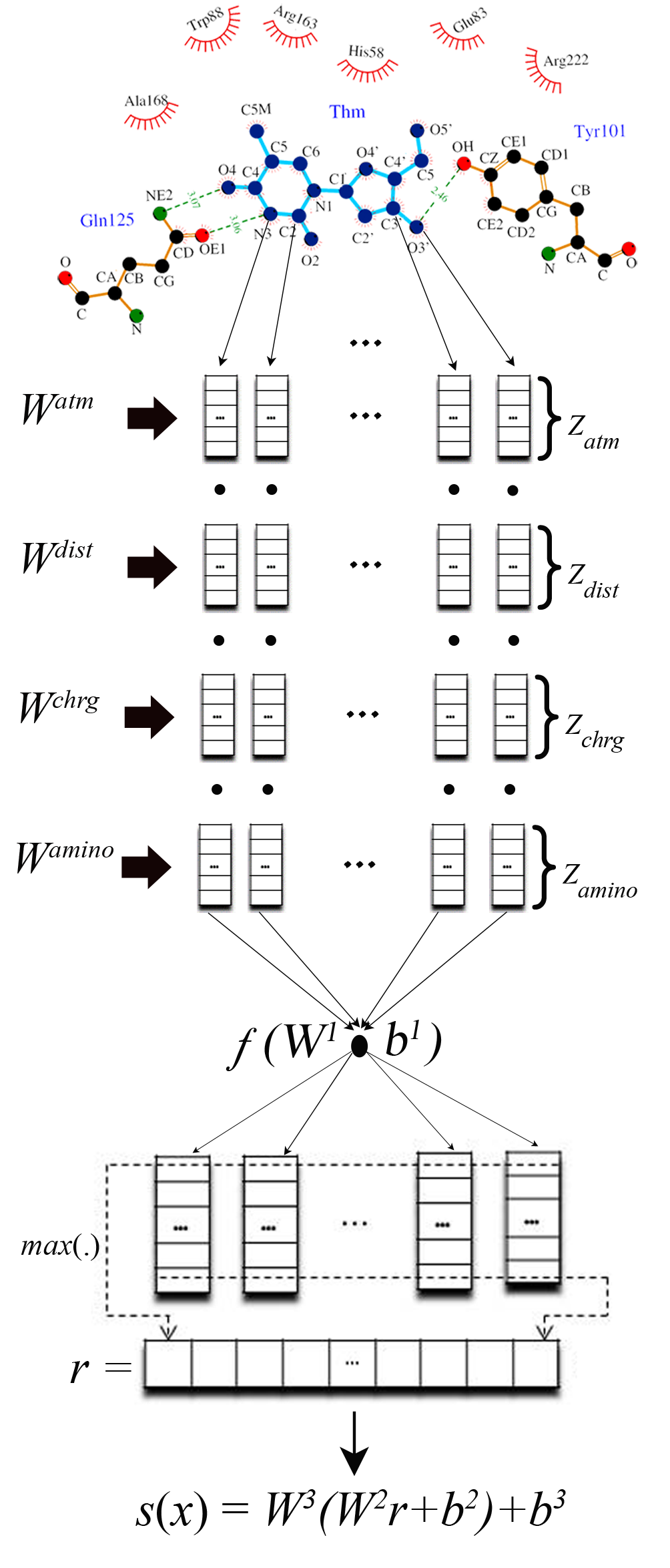}
	\caption{DeepVS architecture scheme. In this figure, we use as an example the compound THM (Thymidine) in complex with  TK - Thymidine kinase protein (ID\_PDB: 1kim) compound atoms are marked in dark blue and their interactions in light blue}
  \label{fgr:figura1}
\end{figure}

DeepVS takes as input the data describing the structure of a protein-compound complex (input layer) and produces a score capable of differentiating ligands from decoys.
Figure~\ref{fgr:figura1} details the DeepVS architecture.
First, given an input protein-compound complex $x$, information is extracted from the local context of each compound atom (First hidden layer). 
The \emph{context} of an atom comprises basic structural data (\emph{basic features}) involving distances, 
neighbor atom types, 
atomic partial charges and associated residues.
Next,
each basic feature from each atom context is converted to feature vectors that are learned by the network.
Then, 
a convolutional layer is employed to summarize information from all contexts from all atoms and generate a distributed vector representation of the protein-compound complex  $r$ (second hidden layer). 
Finally, 
in the last layer (output layer), 
the representation of the complex is given as input to softmax classifier, 
which is responsible for producing the score.
In Algorithm \ref{alg:algorithm1},
we present a high-level pseudo-code with the steps of the feedforward process executed by DeepVS.

\begin{algorithm}[!htb]
\caption{DeepVS feedforeward process}
\label{alg:algorithm1}
\begin{algorithmic}[1]

\State \textbf {Input:} protein-compound complex $x$, where the compound contains $m$ atoms

\State \textbf{Given:} trained network parameters $W^{atm} \in  \mathbb{R}^{d^{atm} \times |A|}$, $W^{dist} \in  \mathbb{R}^{d^{dist} \times |D|}$, $W^{chrg} \in  \mathbb{R}^{d^{chrg} \times |C|}$, $W^{amino} \in  \mathbb{R}^{d^{amino} \times |R|}$,  $W^{1}\in\mathbb{R}^{cf \times \vert{z_{i}}\vert}$, $W^{2}\in\mathbb{R}^{h \times\vert{cf}\vert}$, $W^{3}\in\mathbb{R}^{2 \times\vert{h}\vert}$, $b^{1}\in\mathbb{R}^{cf}$, $b^{2}\in\mathbb{R}^{h}$,
$b^{3}\in\mathbb{R}^{2}$

\State $Z$ = [ ]
\State\textit{// generates the representations of atom contexts}
\For{$i$=1 to $m$}

\State \parbox[t]{\dimexpr\linewidth-\algorithmicindent}{$z_{atm}$ = columns from $W_{atm}$ corresponding to atom types of atom$_i$'s neighbors}
\State \parbox[t]{\dimexpr\linewidth-\algorithmicindent}{$z_{dist}$ = columns from $W_{dist}$ corresponding to distances of atom$_i$'s neighbors}
\State \parbox[t]{\dimexpr\linewidth-\algorithmicindent}{$z_{chrg}$ = columns from $W_{chrg}$ corresponding to charges of atom$_i$'s neighbors}
\State \parbox[t]{\dimexpr\linewidth-\algorithmicindent}{$z_{amino}$ = columns from $W_{amino}$ corresponding to amino acids of atom$_i$'s neighbors}

\State \textit{// representation of the atom$_i$  context}
\State $z_i$ = \{$z_{atm}$; $z_{dist}$; $z_{chrg}$; $z_{amino}$\}
\State $Z$.$add$($z_{i}$)

\EndFor

\State \textit{// U is initialized with zeros}
\State $U$ = $[..] \in  \mathbb{R}^{cf \times m}$
\State \textit{// convolutional layer}
\For{$i$=1 to $m$}
\State $U[:,i] = f(W^{1}Z[i] + b^{1})$
%\State  = u_{i}$
\EndFor

\State \textit{// column-wise max pooling}

\State $r = max(U, axis = 1)$

\State \textit{// hidden and output layers}
\State score = $W^{3}\left(W^{2}r + b^{2}\right) + b^{3}$

\State \textit{// returns normalized score}

\State\Return  $\dfrac{e^{score[1]}}{e^{score[0]}+e^{score[1]}}$

\end{algorithmic}
\end{algorithm}

\subsubsection{Atom context}

First, it is necessary to perform some basic processing on the protein-compound complex to extract the input data for DeepVS. 
The input layer uses information from the context of each atom in the compound. 
The context of an atom ``\emph{a}'' is defined by a set of basic features extracted from its neighborhood.
This neighborhood comprises the $k_{c}$ atoms in the compound closest to ``\emph{a}'' (including itself)
and the $k_{p}$ atoms in the protein that are closest to ``\emph{a}'', 
where $k_{c}$ and $k_{p}$ are hyperparameters that must be defined by the user.
The idea of using information from closest neighbor atoms of both compound and protein have been successfully explored in previous work on structure-based drug design \cite{weber2013vammpire}.

\begin{figure}[!htb]
	\centering
	\includegraphics[scale=0.3]{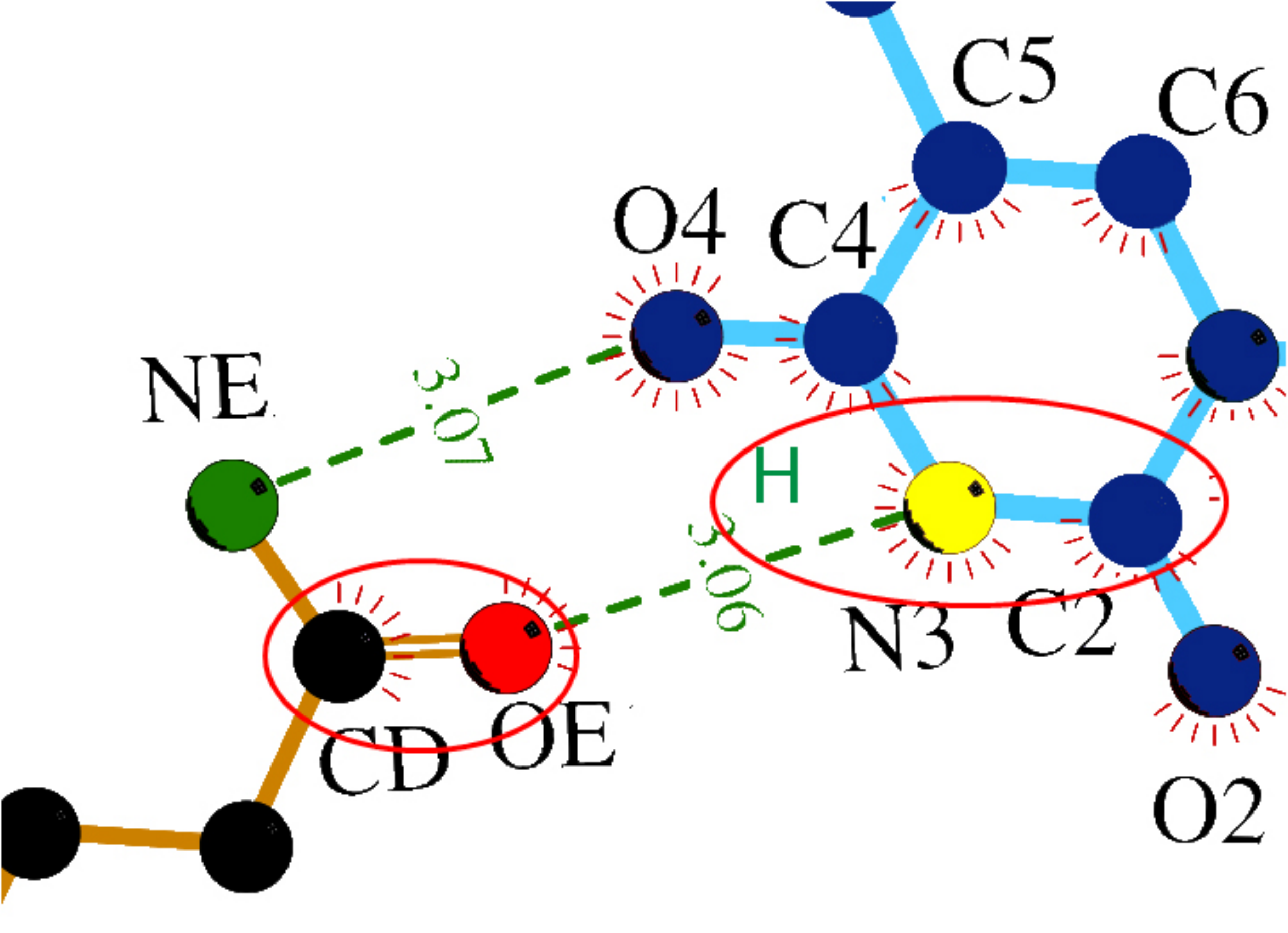}
	\caption{Context from the atom \ce{N3} (yellow) from compound THM (Thymidine). Red circles represent the two closest neighbors from THM compound and the two closest neighbors from the protein closest to \ce{N3} atom.} %Data from \ce{H}-bonds (green dot lines) were not regarded as basic features.
  \label{fgr:figura2}
\end{figure}

The basic features extracted from the context of an atom include the atom types,
atomic partial charges,
amino acid types 
and the distances from neighbors to the reference atom. 
For instance (Figure~\ref{fgr:figura2}), 
for the nitrogen atom (\ce{N3}) from THM compound, 
the vicinity with $k_{c}=3$ 
and $k_{p}=2$ is formed by \ce{N3}, 
\ce{H} and \ce{C} from the compound, 
and the two atoms \ce{OE} and \ce{CD} from the residue \ce{Gln125} in the protein Thymidine kinase (ID\_PDB: 1kim). 
In this particular case, 
the context of the atom \ce{N3} contains the following values for each basic feature: 

\begin{itemize}
%\begin{footnotesize}
\item Atom Type = [\ce{N}; \ce{H}; \ce{C}; \ce{OE}; \ce{CD}]
\item Charge = [-0.24; 0.16; 0.31; -0.61; 0.69]
\item Distance = [0.00; 1.00; 1.34; 3.06; 3.90]
\item Amino Acid Type  = [\ce{Gln}; \ce{Gln}] 
%\end{footnotesize}
\end{itemize}

\subsubsection{Representation of the atom context}

The first hidden layer in DeepVS transforms each basic feature value of atom contexts into real-valued vectors (aka \emph{embeddings}) by a lookup table operation. 
These embeddings contain features that are automatically learned by the network. 
For each type of basic feature, 
there is a corresponding embedding matrix $W$ that stores a column vector for each possible value for that basic feature.
The matrices $W^{atm}$, 
$W^{dist}$, 
$W^{chrg}$ and $W^{amino}$ contain the embeddings of the basic features atom type, 
distance, 
atomic partial charge, 
and amino acid type, 
respectively. 
These matrices constitute the weight matrices of the first hidden layer and are initialized with random numbers before training the network.

\begin{figure}[!htb]
	\centering
	\includegraphics[scale=0.39]{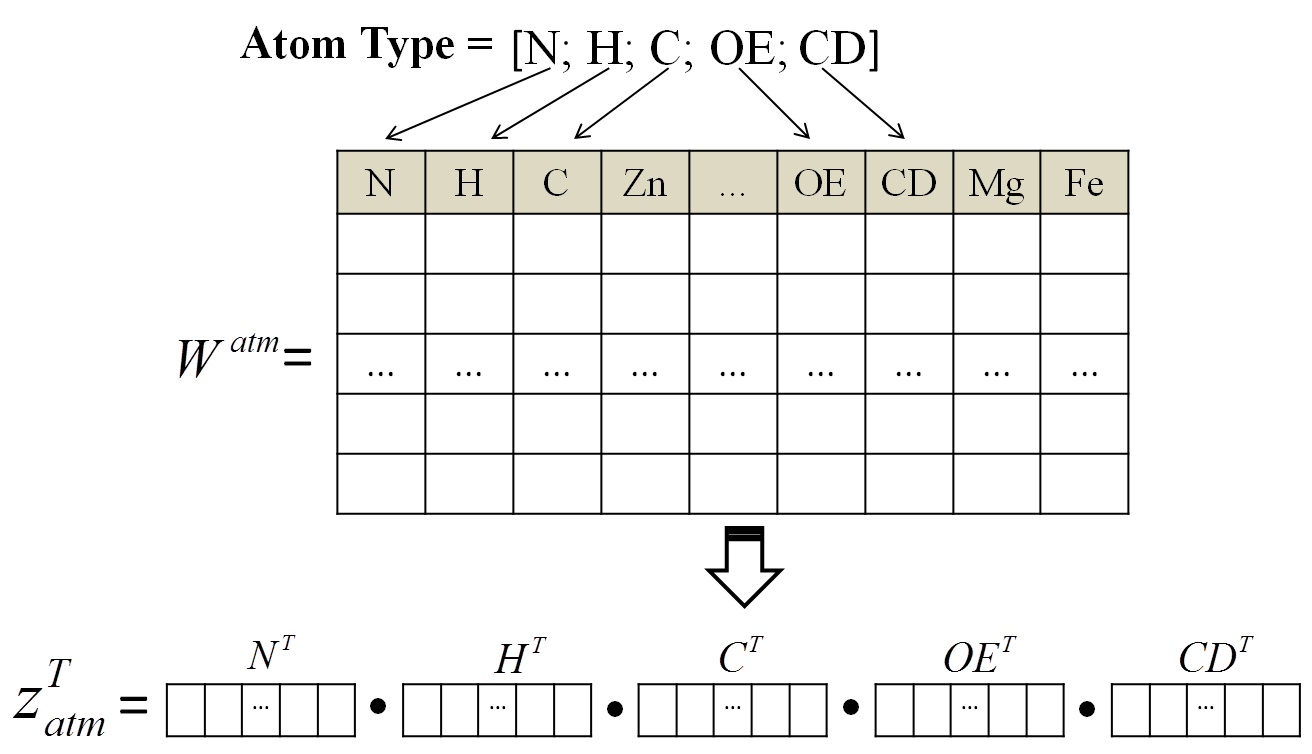}
	\caption{Illustration of the construction of the atom type representation ($z_{atm}$) for the context of the atom \ce{N3} from the compound THM (Thymidine). The symbol \textbullet \ indicates the concatenation operation. }
  \label{fgr:figura3}
\end{figure}

Each column in $W^{atm} \in  \mathbb{R}^{d^{atm} \times |A|}$ corresponds to a feature vector of a particular type of atom, where A is the set of atom types and $d^{atm}$ is the dimensionality of the embedding and constitutes a hyperparameter defined by the user.
Given the context of an atom ``\emph{a}'', 
the network transforms each value of the basic feature atom type in its respective feature vector and then concatenates these vectors to generate the vector atom type representation $z_{atm}$. %(Figure~\ref{fgr:figura3}).
As illustrated in Figure~\ref{fgr:figura3},
retrieving the embedding of an atom type from $W^{atm}$ consists in a simple lookup table operation.
Therefore,
the order of the atom type embeddings (columns) in $W^{atm}$ is arbitrary and have no influence in the result.
However, 
the order in which the embeddings of the atom types are concatenated to form $z_{atm}$ matters.
We always concatenate first the embeddings from atom types of the ligand, from the closest to the farthest, 
then we concatenate the embeddings from atom types of the protein, from the closest to the farthest.

Likewise, 
$z_{dist}$, 
$z_{chrg}$, 
$z_{amino}$ vectors are created from values of distance, 
charge and amino acid types in the context of the target atom. 
Values of the basic features charge and distance need to be discretized before being used as input for the network. 
We define minimum and maximum values, $c_{min}$ and $c_{max}$ respectively, to perform the discretization of charge values. 
Bins equally distanced by 0.05 between minimum and maximum are built. 
For instance, 
with $c_{min}$ = -1 and $c_{max}$ = 1, 
there will be 40 bins.
Similarly, 
to discretize distance values, 
bins equally distributed by 0.3 \r{A} will be defined in the range between $d_{tmin}$ and $d_{tmax}$. 
For example, 
with $d_{tmin}$ = 0 e $d_{tmax}$ = 5.1 \r{A}, 
there will be 18 bins. 

Finally, 
the representation of the context of the atom ``\emph{a}'' is defined as $z_a=\{z_{atm}; z_{dist}; z_{chrg}; z_{amino}\}$,
comprising the concatenation of the vectors previously described.
Our hypothesis is that from the basic contextual features, 
the network can learn more abstract features (the \emph{embeddings}) that are informative about the discrimination between compound and decoys. 
This type of strategy, 
where basic features (words) are transformed into more abstract features (word embeddings), 
has obtained enormous success in the Natural Language Processing (LNP) field \cite{Collobert2011,Socher2012,word2vec2013,DosSantos2014,DosSantosb2014}.

\subsubsection{Representation of the protein-compound complex}

The second hidden layer in DeepVS is a convolutional layer, responsible for (1) extracting more abstract features from the representations of all atom contexts in the compound, 
and (2) summarizing this information in a fixed-length vector $r$.
We name the vector $r$ the \emph{representation of the compound-protein complex},
and it is the output of the convolutional layer.

The subjacent goal in using a convolutional layer is its ability to deal with inputs of variable sizes \cite{Waibel1989}.
In the case of virtual screening, 
different compounds can have a different number of atoms.
Therefore, 
the number of representations of atom contexts can differ for different complexes.
In DeepVS,
the convolutional layer allows the processing of complexes of different sizes.

Given a complex $x$, 
whose compound is composed of $m$ atoms, 
the input to the convolutional layer is a list of vectors $\{z_{1},z_{2},\cdots,z_{m}\}$, 
where $z_{i}$ is the representation of the context of the $i$-th atom in the compound. 
In the first stage of the convolutional layer, generation of more abstract features from each vector $z_{i}$ is carried out according to:
\begin{equation}
  u_i =f(W^{1}z_{i} + b^1)        
	\label{eqn:equation1}
\end{equation}
where $W^{1}\in\mathbb{R}^{cf \times \vert{z_{i}}\vert}$ is the weight matrix corresponding to the convolutional layer, 
$b^{1}$ is a bias term,
$f$ is the hyperbolic tangent function and $u{_i}\in\mathbb{R}^{cf}$ corresponds to the resulting feature vector. 
The number of units (also called filters) in the convolutional layer, $cf$, is a hyperparameter defined by the user. 

The second stage in the convolutional layer,
also known as pooling layer, 
summarizes the features from the various atom contexts. 
The input consists of a set of vectors $\{u_{1},u_{2},\cdots,u_{m}\}$. For the DeepVS, we use a max-pooling layer, 
which produces a vector $r\in\mathbb{R}^{cf}$, where the value of the $j-th$ element is defined as the maximum of the $j-th$ elements of the set of input vectors, {\itshape i.e.}:
\begin{equation}
  [r]_{j}=\max_{1 \le i \le m}⁡[u_{i}]_{j}        
	\label{eqn:equation2}
\end{equation}

The resulting vector $r$ from this stage is the representation of the compound-protein complex (Eq.~\ref{eqn:equation2}). 
In this way, the network can learn to generate a vector representation that summarizes the information from the complex that is relevant to discriminate between ligands and decoys. 

\subsubsection{Scoring of Compound-Protein Complex}
The vector $r$ is processed by two usual neural network layers:
a third hidden layer that extract one more level of representation,
and an output layer, 
which computes a score for each one of the two possible classifications of the complex: 
(0) inactive compound and (1) active compound. 
Formally, 
given the representation $r$ generated for the complex $x$, the third hidden layer and the output layer execute the following operation:
\begin{equation}
  s(x)=W^{3}\left(W^{2}r+b^{2}\right)+b^{3}
	\label{eqn:equation3}
\end{equation}
where 
$W^{2}\in\mathbb{R}^{h \times\vert{cf}\vert}$ is the weight matrix of the third hidden layer,
$W^{3}\in\mathbb{R}^{2 \times\vert{h}\vert}$ is the weight matrix of the output layer, 
$b^{2}\in\mathbb{R}^{h}$ and $b^{3}\in\mathbb{R}^{2}$ are bias terms.
The number of units in the hidden layer, $h$, is a hyperparameter defined by the user.
$s(x)\in\mathbb{R}^{2}$ is a vector containing the score for each of the two classes.

Let $s(x)_{0}$ and $s(x)_{1}$ be the scores for the classes 0 and 1, respectively.
We transform these scores in a probability distribution using the \emph{softmax} function, as follows: 
\begin{equation}
  p(0|x)=\frac{e^{s(x)_{0} }}{e^{s(x)_{0}}+e^{s(x)_{1}}}            
	\label{eqn:equation4}
\end{equation}

\begin{equation}
  p(1|x)=\frac{e^{s(x)_{1} }}{e^{s(x)_{0}}+e^{s(x)_{1}}}            
	\label{eqn:equation5}
\end{equation}
where we interpret $p(0|x)$ and $p(1|x)$ as the conditional probabilities of the compound be a decoy or a ligand, respectively,
given the compound-protein complex data acquired from docking.

The likelihood of class 1 (active ligand) is the scoring used to rank ligands during Virtual Screening essays. 
The larger the scoring, 
the greater the chance the compound is an active ligand.

\subsubsection{Training DeepVS}

The common approach for training neural networks is the stochastic gradient descent (SGD) algorithm \cite{Rumelhart1988}.
In our case,
SGD is used to minimize a loss function over a training set $D$ that contains both complexes of ligands and decoys. 
At each iteration, 
a new complex $(x,y)\in D$ is randomly chosen, 
where $y=1$ if the complex contains an active ligand and $y=0$, 
otherwise.
Next, 
the DeepVS network with parameter set $\theta=\{W^{atm}$, $W^{chrg}$, $W^{dist}$, $W^{amino}$, $W^{1}$, $b^{1}$, $W^{2}$, $b^{2}$, $W^{3}$, $b^{3}\}$ is used to estimate the probability $p(y|x,\theta)$. 
Finally, the prediction error is computed as the negative log-likelihood, $-\log{(p(y|x,\theta))}$, 
and the network parameters are updated applying the backpropagation algorithm \cite{Rumelhart1988}. 
In other words, 
the set of parameters $\theta$ of the network is learned by using SGD to select a set of values that minimize the loss function with respect to $\theta$:
\begin{equation}
 \theta \longmapsto \sum_{(x,y)\in D}-\log⁡  {p(y|x,\theta)}            
	\label{eqn:equation6}
\end{equation}

In our experiments, 
we applied SGD with minibatches,
which means that instead of considering only one compound-protein complex at each iteration
we consider a small set of $m_s$ randomly selected complexes and used the average prediction loss to perform backpropagation.
We set $m_s=20$ in our cases.
Also, we used Theano \cite{Bergstra2010} to implement DeepVS and perform all the experiments reported in this work.

\subsection{Experimental Setup}

\subsubsection{Dataset}

We used the Directory of Useful Decoys (DUD) \cite{Huang2006} as a benchmark to evaluate our deep-learning-based virtual screening approach. 
One of the main reasons to use this dataset was the  possibility of comparing our results with the ones from systems previously proposed for revalidation of scoring functions and reclassification in virtual screening \cite{Durrant2011,Durrant2013,Arciniega2014}.
There is a problem in the partial charges of the original version of DUD that makes it trivial to discriminate between ligands and decoys.
Therefore, 
we use the version of DUD produced by Armstrong et al.\cite{Armstrong2010}, which contains corrected atomic partial charges.

The DUD dataset has been developed specifically to validate VS methods in a rigorous way. 
The dataset is composed of 40 receptors distributed in six different biological groups: 
Hormone nuclear receptors, 
kinases, 
serine proteases, 
metalloenzymes, 
flavoenzymes and other classes of enzymes \cite{Huang2006}. Also, 
it possesses 2,950 annotated ligands and 95,316 decoys, a ratio of 36 decoys for each annotated ligand. 
Each of the 36 decoys was retrieved from the ZINC databank, to mimic some physical property of the associated ligand, such as molecular weight, 
cLogP, 
and the number of H-bonds groups, 
although differing in its topology \cite{Arciniega2014,Huang2006}.

\subsubsection{Docking Programs}

In this work, 
we used two different computer programs to perform molecular docking: Dock 6.6 \cite{Lang2009} and AutodockVina1.1.2 \cite{Trott2010}. 
Both are open access and widely used in the academia to perform VS. 
Dock 6.6 offers physics-based energy score functions based on force fields and scoring functions (GRID score \& AMBER score) \cite{Lang2009}. 
Autodockvina1.1.2 applies a hybrid scoring function combining characteristics of knowledge-based and empiric scoring functions (Vina score) \cite{Trott2010}.

\subsubsection{Dock 6.6 setup}
Protein and compound structures were prepared using computational tools from Chimera \cite{Pettersen2004}. Receptors were prepared using the DockPrep module from Chimera. Ligands, non-structural ions, solvent molecules, and cofactors were removed. Absent atoms were added, and Gasteiger atomic partial charges were applied. Input files were .mol2 formatted, except those receptors that were used to calculate molecular surface, 
in which Hydrogen atoms were removed, and were finally saved in pdb format \cite{Pettersen2004}.  
Spheres were created with radius in the range 13.5-15.5 \AA, varying according to the size of the active site of the receptor. 
Box and grid parameters were taken directly from DUD. The docking procedure was performed according to an available script provided by Dock6.6 program \cite{Lang2009}.

\subsubsection{AutodockVina1.1.2 Setup}
Receptors and compound were prepared following default protocols and Gasteiger atomic partial charges were applied \cite{Trott2010,Morris2009}. 
A cubic Grid of edge as 27 $\AA$ was defined. 
The center of the grid box coincided with the center of mass of the ligand. 
Docking runs were performed following the default settings defined in AutoDockTools \cite{Morris2009}.
The only hyperparameter we changed was the global search exhaustiveness,
which we set to 16 as in Arciniega \& Lange (2014) \cite{Morris2009}.
It is worth to note that, 
although AutodockVina1.1.2 can output more than one pose,
in our experiments, we only consider just one,
which corresponded to the pose that AutodockVina1.1.2 outputs as the best one.

\subsubsection{Evaluation Approach}
\begin{figure}[!htb]
	\includegraphics[scale=1]{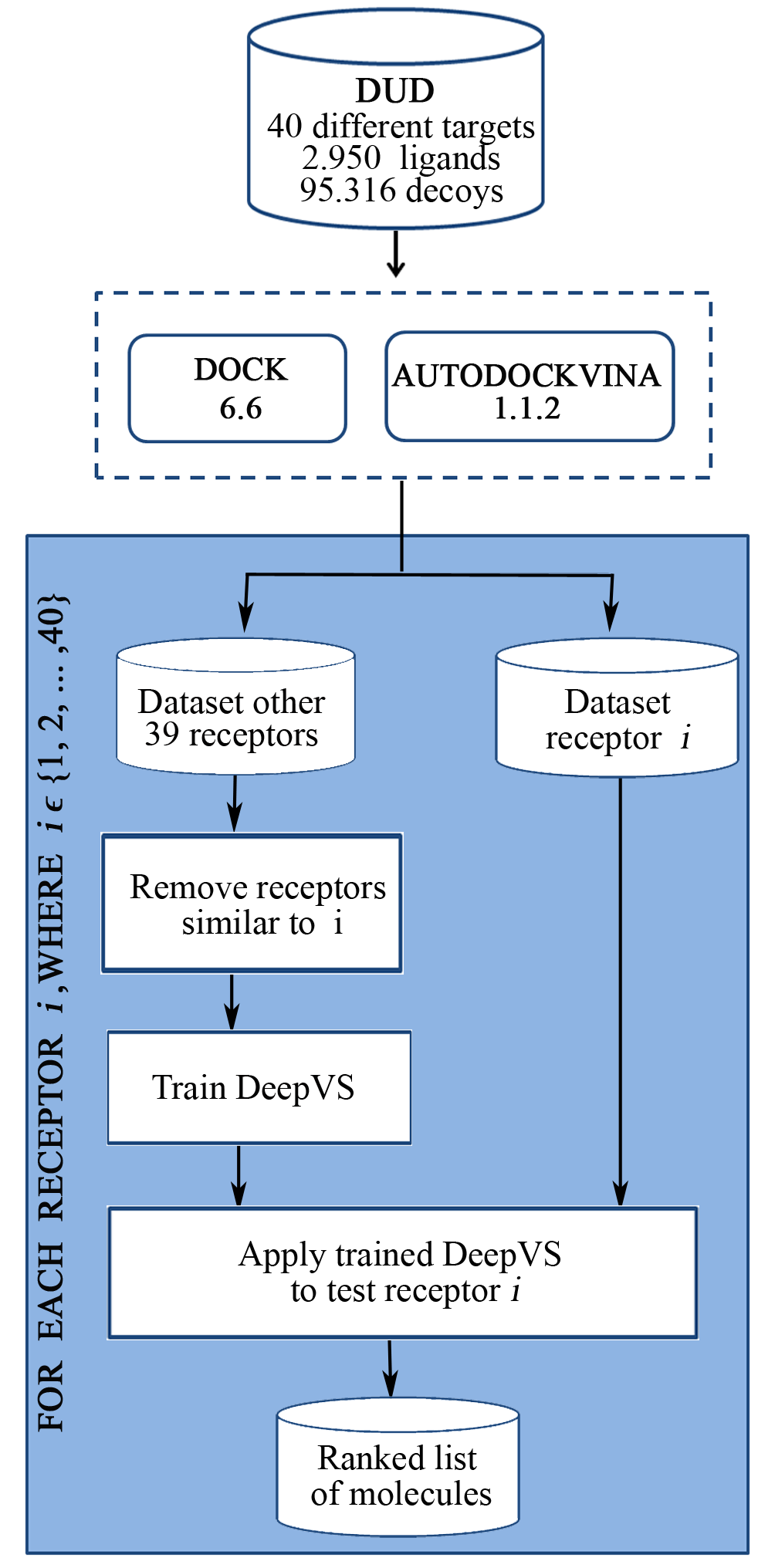}
  \caption{Leave-one-out cross-validation.}
  \label{fgr:figura4}
\end{figure}

The performance of the proposed method is assessed using leave-one-out cross-validation 
with the 40 proteins from the DUD dataset.
Figure~\ref{fgr:figura4} illustrates the process we follow to perform our leave-one-out cross-validation experiments.
First, we applied either DOCK6.6 or AutodockVina1.1.2 for each protein and its respective set of ligands and decoys to generate the docked protein-compound complexes.
Next,
in each run, 
one receptor was left out of the test set while the others were employed as the training set.

To avoid distortions in the performance results of DeepVS, it was essential to remove all receptors similar to the one used as a test in a specific cross-validation iteration from the training set.
Following Arciniega \& Lange (2014) \cite{Arciniega2014},
we regarded as similar receptors those sharing the same biological class or those with reported positive cross enrichment \cite{Huang2006}. 
Once the network was trained, 
it was applied to the test receptor, 
producing a scoring for each of the potential ligands. Such score was used to rank the set compounds. 
The ranking was validated using metrics that indicate the algorithm's performance.

\subsubsection{DeepVS Hyperparameters}

The main advantage in using leave-one-out cross-validation is the possibility of tuning the neural network hyperparameters without being much concerned with overfitting.
In fact,
leave-one-out cross-validation is a suitable method for tuning hyperparameters of machine learning algorithms when the dataset is small \cite{Arlot2010}.
In our experiments,
we used the same set of hyperparameters for the 40 leave-one-out cross-validation iterations.
This is equivalent to perform 40 different experiments with different training/test sets using the same configuration for DeepVS.  
The hyper-parameter values that provided our best results and were used in our experiments with both, AutodockVina1.1.2 and Dock6.6, are specified in Table~\ref{tbl:Table1}.
Note that our evaluation approach was stricter than the one used by Arciniega \& Lange (2014) \cite{Arciniega2014}, 
because they tuned the hyper-parameters using a hold-out set at each leave-one-out iteration.

In the next section, we present some experimental results that detail the difference in performance when we vary some of the main hyper-parameters of DeepVS.

\begin{table}[!htb]\footnotesize
  \caption{Hyperparameter values for DeepVS used to train the network.}
  \label{tbl:Table1} 
   \setlength{\tabcolsep}{3pt}
   \begin{tabular}{lll}
		\hline
         \multicolumn{1}{c}{ Hyperparameter} & \multicolumn{1}{c}{ Description} & \multicolumn{1}{c}{ Value}  \\
		\hline
		 $d^{atm}$   & { Atom type embedding size}    & 200   \\
         $d^{amino}$ & { Amino Acid emb. size}        & 200   \\
         $d^{chrg}$  & { Charge emb. size}            & 200   \\
         $d^{dist}$  & { Distance emb. size}          & 200   \\
	     $cf$        & { \# conv. filters}            & 400   \\
	     $h$         & { \# hidden units}             & 50   \\
	     $\lambda$   & { Learning rate}               & 0.075 \\
	     $k_{c}$     & { \# neig. atoms from comp.}   & 6  \\
	     $k_{p}$     & { \# neig. atoms from protein} & 2  \\
       
       \hline
  \end{tabular}
\end{table}

\subsection{Evaluation Metrics}

To validate DeepVS performance and compare it with other methods previously published in the literature,
we used two well\-established VS performance metrics: 
the enrichment factor (EF) and the area under the ROC curve (Receiver Operating Characteristic) \cite{Jahn2011,Nicholls2008}.

ROC curves are a way to represent the relationship between the selectivity (Se) and specificity (Sp) along a range of continuous values (Equations~\ref{eqn:equation7} and \ref{eqn:equation8}). 
It represents the ratio of true positives in function of the false-positives.

\small
\begin{equation}
Se = \frac{true \ positives}{total \ actives}
\label{eqn:equation7}
\end{equation}
\normalsize 

\small
\begin{equation}
Sp = \frac{true \ negatives}{total \ decoys}       
\label{eqn:equation8}
\end{equation}
\normalsize 

The area under the ROC curve (AUC) represents a quantification of the curve and facilitates the comparison of results. 
The AUC is calculated as given in Eq.~\ref{eqn:equation9}, where $N_{actives}$ depicts the number of actives, $N_{decoys}$ represents the number of decoys, 
and $N^{i}_{decoys\_seen}$ describes the number of decoys that are higher ranked than the $i$-th active structure  \cite{Jahn2011}.
An AUC $\leqslant$ 0.50 indicates a random selection,
whereas an AUC of 1.0 indicates the perfect identification of active compounds.

\begin{equation}
AUC = 1-\frac{1}{N_{actives}}\sum_{i}^{N_{actives}}\frac{N^{i}_{decoys\_seen}}{N_{decoys}}
\label{eqn:equation9}
\end{equation}

Given the set of compounds ranked by score,
the enrichment factor at x\% (Eq.~\ref{eqn:equation10}) informs how good is the set formed by the top x\% ranked compounds compared to a set of an equal size selected at random from the entire set of compounds\cite{Jahn2011,Nicholls2008}.
The EF is computed as

\small
\begin{equation}
EF_{x\%} = \frac{actives \ at \  X\%}{ compounds \ at \  X\%} \frac{total \  compounds}{total \  actives}
\label{eqn:equation10}
\end{equation}
\normalsize

\section{Results and Discussion} 
\label{sec:results}

\subsection{DeepVS vs. Docking Programs}

\begin{table*}[!htb]\scriptsize
  \caption{Virtual screening AUC ROC and enrichment factor (\emph{ef}) results for Dock6.6, AutodockVina1.1.2 and DeepVS.}
  \label{tbl:deepvs_vs_docking} 
	\setlength{\tabcolsep}{4pt}
	\begin{tabular}{lrrrrrrrrrrrrrrrrrrr}
		\hline
		 &\multicolumn{4}{c}{Dock}& & \multicolumn{4}{c}{DeepVS-Dock}& & \multicolumn{4}{c}{ADV}& & \multicolumn{4}{c}{DeepVS-ADV}\\
         \cline{2-5} \cline{7-10} \cline{12-15} \cline{17-20}
         &{\tiny $ef_{max}$}&{\tiny $ef_{2\%}$}&{\tiny $ef_{20\%}$}& auc & &{\tiny $ef_{max}$}&{\tiny $ef_{2\%}$}&{\tiny $ef_{20\%}$}&auc& &{\tiny $ef_{max}$}&{\tiny $ef_{2\%}$}&{\tiny $ef_{20\%}$}&auc& &{\tiny $ef_{max}$}&{\tiny $ef_{2\%}$}&{\tiny $ef_{20\%}$}&auc \\
	\hline
		 average         & 20.1 &  5.3 & 1.3 & 0.48 & & 16.9 & 5.9 & 3.0 & 0.74 & & 16.4 &  6.0 & 2.0 & 0.62 & & 16.0 & 6.6 & 3.1 & \bf 0.81\\
		\hline
		 ACE             & 36.2 &  3.0 & 0.6 & 0.41 & & 3.6 & 2.0 & 2.7 & \bf 0.75 & &  9.1 &  3.0 & 1.4 & 0.38 & & 2.2 & 1.0 & 1.8 & 0.72\\
         AChE            & 12.1 &  0.5 & 0.8 & 0.50 & & 1.6 & 0.5 & 0.8 & 0.48 & &  5.4 &  4.8 & 3.0 & \bf 0.68 & & 4.0 & 1.9 & 1.1 & 0.51\\
         ADA             & 24.5 & 13.0 & 3.3 &  0.86  & & 8.5 & 6.5 & 4.1 & \bf 0.87 & &  1.1 &  0.0 & 1.1 & 0.47 & & 9.2 & 2.2 & 3.3 & 0.83\\
         ALR2            &  1.0 &  0.0 & 0.2 & 0.38 & & 1.7 & 0.0 & 1.3 & 0.56 & &  8.2 &  3.8 & 2.5 & \bf 0.70 & & 5.7 & 3.8 & 1.3 & 0.66\\
		 AmpC            &  5.1 &  4.8 & 1.4 & \bf 0.57 & & 1.3 & 0.0 & 0.0 & 0.44 & &  1.0 &  0.0 & 0.2 & 0.23 & & 1.1 & 0.0 & 0.0 & 0.42\\
		 AR              &  7.3 &  0.7 & 0.1 & 0.25 & & 36.5 & 18.3 & 4.1 & 0.80 & & 36.5 & 14.2 & 3.5 & 0.74 & & 21.9 & 11.5 & 4.2 & \bf 0.88 \\
		 CDK2            & 36.6 & 11.9 & 2.1 & 0.56 & & 11.4 & 8.9 & 3.7 & \bf 0.82 & & 18.3 &  8.9 & 2.1 & 0.66 & & 6.1 & 4.0 & 2.9 & 0.79\\
		 COMT            & 20.0 & 17.8 & 2.7 & 0.64 & & 40.1 & 13.4 & 3.2 & 0.88 & & 40.1 &  8.9 & 1.4 & 0.41 & & 20.0 & 8.9 & 4.6 & \bf 0.92\\
		 COX\-1          &  1.1 &  0.0 & 0.2 & 0.35 & & 35.0 & 8.2 & 3.0 & 0.75 & & 20.0 & 10.3 & 3.6 & \bf 0.79 & & 5.0 & 2.1 & 2.8 & 0.77 \\
		 COX\-2          &  1.3 &  0.7 & 1.2 & 0.58 & & 18.4 & 12.4 & 3.6 & 0.78 & & 36.8 & 21.4 & 3.8 & 0.84 & & 36.8 & 12.7 & 4.3 & \bf 0.91 \\ 
		 DHFR            & 36.5 &  3.7 & 1.3 & 0.48 & & 18.3 & 10.4 & 4.5 & 0.88 & & 36.5 &  8.2 & 3.5 & 0.86 & & 9.1 & 6.7 & 4.8 & \bf 0.94\\
		 EGFr            & 34.5 & 12.0 & 1.9 & 0.49 & & 23.0 & 8.4 & 4.6 & \bf 0.93 & &  4.9 &  2.5 & 1.5 & 0.58 & & 8.6 & 5.5 & 3.6 &  0.86\\
		 ER$_{agonist}$    & 18.1 &  2.3 & 1.0 & 0.43 & & 6.5 & 0.8 & 3.5 & 0.75 & & 17.7 & 16.6 & 3.3 & 0.79 & & 8.1 & 6.0 & 3.9 & \bf 0.88 \\
		 ER$_{antagonist}$ &  7.1 &  5.9 & 2.2 & 0.68 & & 7.1 & 5.9 & 4.0 & \bf 0.90  & & 13.8 & 8.9 & 2.3 & 0.66 & & 7.4 & 3.8 & 3.8 & 0.88\\
		 FGFrl           & 36.6 &  8.9 & 2.1 & 0.49 & & 15.7 & 8.1 & 4.6 & \bf 0.91 & &  1.1 &  0.0 & 1.0 & 0.48 & & 36.6 & 7.7 & 3.3 & 0.85\\
		 FXa             & 36.9 &  3.2 & 1.1 & 0.49 & & 2.0 & 0.4 & 1.7 & 0.71 & &  3.2 &  2.1 & 1.5 & 0.66 & & 4.3 & 1.4 & 3.9 & \bf 0.86\\
		 GART            & 10.5 &  7.4 & 4.5 &  0.90 & & 12.3 & 4.9 & 4.8 & \bf 0.92 & &  2.9 &  0.0 & 2.6 & 0.77 & & 2.6 & 0.0 & 2.4 & 0.77\\
		 GPB             &  1.0 &  0.0 & 0.1 & 0.23 & & 2.7 & 1.9 & 1.2 & 0.51 & &  3.1 &  2.9 & 1.2 & \bf 0.52 & & 1.0 & 0.0 & 0.9 & 0.42\\
		 GR              & 18.4 &  1.9 & 0.3 & 0.22 & & 11.1 & 7.6 & 2.2 & 0.49 & & 18.4 &  4.4 & 1.2 & 0.57 & & 20.3 & 10.8 & 4.4 & \bf 0.91\\
		 HIVPR           &  1.0 &  0.0 & 0.4 & 0.16 & & 4.1 & 0.9 & 2.1 & 0.51 & & 36.6 &  6.6 & 2.6 & 0.72 & & 6.2 & 5.6 & 4.1 & \bf 0.88\\
		 HIVRT           & 36.9 &  4.9 & 1.0 & 0.40 & & 36.9 & 6.2 & 2.8 & 0.69 & & 24.6 &  6.2 & 1.8 & 0.64 & & 7.4 & 4.9 & 2.3 & \bf 0.73\\
		 HMGR            & 18.2 &  4.2 & 0.4 & 0.16 & & 36.5 & 2.8 & 1.0 & 0.24 & &  1.0 &  0.0 & 0.6 & 0.42 & & 36.5 & 19.6 & 4.9 & \bf 0.96 \\
		 HSP90           &  2.3 &  0.0 & 1.0 & 0.39 & & 5.8 & 2.0 & 3.3 & 0.74 & &  1.3 &  0.0 & 0.8 & 0.54 & & 10.0 & 4.1 & 5.0 & \bf 0.94\\
		 InhA            & 36.7 &  5.3 & 1.2 & 0.38 & & 36.7 & 13.0 & 4.3 & 0.90 & & 36.7 & 11.2 & 1.8 & 0.54 & & 6.7 & 16.6 & 4.5 & \bf 0.94\\
		 MR              & 18.3 &  3.3 & 1.0 & 0.36 & & 14.7 & 6.7 & 2.3 & 0.55 & & 36.7 & 20.0 & 4.0 & \bf 0.82 & & 24.4 & 16.7 & 3.3 & \bf 0.82 \\
		 NA              & 36.6 & 13.2 & 3.3 & 0.73 & & 3.5 & 0.0 & 2.8 & \bf 0.78 & &  1.1 &  0.0 & 0.5 & 0.40 & & 2.1 & 0.0 & 1.0 & 0.68\\
		 P38MAP          & 33.8 & 11.1 & 2.2 & 0.62 & & 33.8 & 16.0 & 4.2 & \bf 0.91 & &  2.8 &  1.4 & 2.2 & 0.62 & & 21.6 & 16.4 & 3.9 & 0.87 \\
		 PARP            & 36.6 & 10.7 & 1.5 & 0.51 & & 2.9 & 0.0 & 0.8 & 0.63 & & 36.6 &  4.6 & 2.9 & \bf 0.74 & & 1.7 & 0.0 & 0.8 & 0.65 \\
		 PDE5            & 36.5 &  3.0 & 0.9 & 0.39 & & 12.2 & 3.9 & 2.9 & 0.75 & & 36.5 &  6.9 & 1.7 & 0.57 & & 36.5 & 8.9 & 3.1 & \bf 0.86\\
		 PDGFrb          & 36.8 &  5.4 & 1.2 & 0.38 & & 17.4 & 14.4 & 4.5 & \bf 0.92 & & 36.8 &  5.4 & 1.1 & 0.48 & & 36.8 & 19.2 & 3.7 & 0.91\\
		 PNP             &  4.8 &  4.0 & 0.6 & 0.44 & & 10.3 & 2.0 & 4.8 & \bf 0.94 & &  3.0 &  2.0 & 1.8 & 0.66 & & 4.2 & 0.0 & 4.0 & 0.86\\
		 PPARg           &  2.6 &  1.8 & 0.4 & 0.49 & & 36.9 & 1.2 & 2.0 & 0.79 & &  4.1 &  3.1 & 1.7 & 0.64 & & 36.9 & 2.5 & 4.4 & \bf 0.87\\
		 PR              &  3.1 &  1.8 & 0.6 & 0.33 & & 4.8 & 1.8 & 3.3 & 0.70 & &  1.5 &  0.0 & 1.1 & 0.43 & & 8.5 & 5.5 & 2.4 & \bf 0.77\\
		 RXRa            & 36.4 & 12.1 & 2.0 & 0.73 & & 6.4 & 4.9 & 4.2 & 0.91 & & 36.4 & 24.3 & 4.2 & \bf 0.93 & & 12.1 & 9.7 & 3.0 & 0.85\\
		 SAHH            &  1.7 &  1.5 & 1.1 & 0.50 & & 19.3 & 15.1 & 4.7 & 0.94 & & 13.5 & 10.5 & 3.2 & 0.80 & & 19.9 & 18.1 & 4.7 & \bf 0.95\\
		 SRC             & 38.4 & 12.3 & 1.8 & 0.44 & & 25.6 & 12.9 & 3.9 & \bf 0.88 & &  4.0 &  2.9 & 2.2 & 0.69 & & 19.2 & 9.7 & 3.4 & 0.85\\
		 thrombin        &  3.6 &  2.3 & 1.4 & 0.60 & & 36.3 & 5.4 & 1.1 & 0.59 & & 18.1 &  6.2 & 2.8 & 0.73 & & 36.3 & 6.2 & 3.2 & \bf 0.83\\
		 TK              &  2.0 &  0.0 & 1.4 & \bf 0.63 & & 1.2 & 0.0 & 0.5 & 0.44 & &  1.5 &  0.0 & 0.2 & 0.55 & & 1.4 & 0.0 & 0.2 & 0.54\\
		 trypsin         & 36.1 &  4.5 & 1.6 & 0.51 & & 18.0 & 5.6 & 1.9 & 0.65 & &  9.8 &  3.4 & 1.6 & 0.63 & & 36.1 & 6.8 & 2.4 & \bf 0.80\\
		 VEGFr2          & 36.7 & 10.9 & 1.4 & 0.43 & & 9.7 & 6.1 & 3.8 & 0.88 & & 36.7 &  5.4 & 1.6 & 0.56 & & 36.7 & 4.8 & 4.1 & \bf 0.90\\
		 
    \hline
  \end{tabular}
\end{table*}

In Table~\ref{tbl:deepvs_vs_docking} we report,
for each of the 40 receptors in DUD,
the virtual screening performance for Dock6.6,
AutodockVina1.1.2 (henceforth ADV),
DeepVS using Dock6.6 output (DeepVS-Dock) 
and DeepVS using AVD output (DeepVS-ADV).
For each system,
we report the AUC ROC and the enrichment factor ($ef$) at 2\% and 20\%.
We also report the $ef$ maximum ($ef_{max}$),
which is the maximum value of $ef$ that can be achieved for a given ranked list of compounds.
Among the four systems,
DeepVS-ADV achieved the best average result for AUC,
$ef_{2\%}$ and $ef_{20\%}$.
DeepVS-ADV had the best AUC for 20 out of the 40 DUD receptors.

Overall,
the quality of the docking output impacts the performance of DeepVS,
which is expected.
The average performance of DeepVS-ADV,
which had a better docking input from ADV (avrg. AUC of 0.62)
produced better AUC, $ef_{2\%}$, $ef_{20\%}$ and $ef_{max}$ than DeepVS-Dock,
whose input is based on DOCK6.6 (avrg. AUC of 0.48).
On the other hand,
there were some cases where the AUC of docking program was very poor,
but DeepVS was able to boost the AUC result significantly.
For instance,
although DOCK6.6 produced an AUC < 0.40 for the receptors AR, COX1, HSP90, InhA, PDE5, PDGFrb and PR,
DeepVS-Dock resulted in an AUC > 0.70 for these receptors.

\begin{figure}[!htb]
	\includegraphics[scale=0.4]{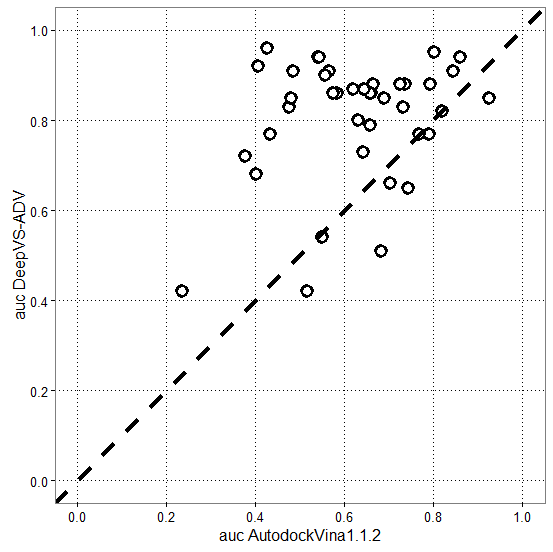}
	\caption{DeepVS-ADV vs AutodockVina1.1.2 AUC. The circles represent each of the 40 DUD receptors.} 
  \label{fgr:deepvs_vs_vina}
\end{figure}

In Figure \ref{fgr:deepvs_vs_vina} we compare the AUC of DeepVS-ADV and ADV for the 40 receptors.
DeepVS-ADV achieved AUC > 0.70 for 33 receptors, 
while this number was only 13 for ADV.
The number of receptors with AUC $<$ 0.50 was 2 for DeepVS-ADV and 9 for ADV.
The AUC of DeepVS-ADV was higher than the one for ADV for 31 receptors.
In average,
the AUC of DeepVS-ADV (0.81) was 31\% better than the one for ADV (0.62).
Additionally,
when we selected 20\% of the data according to the ranked compounds,
in average DeepVS-ADV's $ef$ (3.1) was 55\% larger than the $ef$ from the ADV (2.0).

\begin{figure}[!htb]
	\includegraphics[scale=0.4]{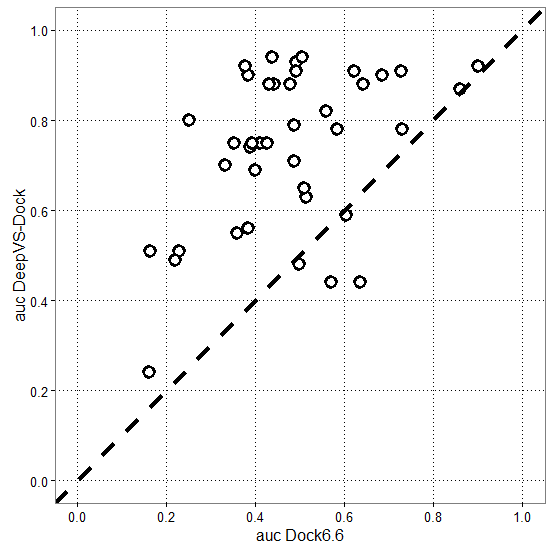}
	\caption{DeepVS-Dock vs DOCK6.6 AUC. The circles represent each of the 40 DUD receptors.} 
  \label{fgr:deepvs_vs_dock6}
\end{figure}

%%% compare this in percentage 
A comparison of the AUC from DeepVS-Dock and Dock6.6 for the 40 receptors is presented in Figure \ref{fgr:deepvs_vs_dock6}.
In average,
the AUC of DeepVS-Dock (0.74) was 54\% better than the one for the Dock6.6 (0.48).
While Dock6.6 achieved AUC > 0.70 for 10\% (4) of the receptors only,
DeepVS-Dock reached AUC > 0.70 for 68\% (27) of the receptors.
The number of receptors with AUC $<$ 0.50 was 5 for DeepVS-Dock and 23 for Dock6.6.
The AUC of DeepVS-Dock was higher than the one for Dock6.6 for 36 receptors.
Finally,
when we select 20\% of the data according to the ranked compounds,
in average DeepVS-Dock's $ef$ (3.0) was more than two times larger than the $ef$ from Dock6.6 (1.3).

The experimental results presented in this section,
which include outputs from two different docking programs, 
are strong evidence that DeepVS can be an effective approach for improving docking-based virtual screening.

\subsection{DeepVS Sensitivity to Hyperparameters}
\label{sec:deepvs_sensitivity}

In this section, 
we present experimental results regarding an investigation on the sensitivity of DeepVS concerning the main hyper-parameters.
In our experiments,
we used the ADV output as input to DeepVS.
Therefore,
all results reported in this section were generated using DeepVS-ADV.
However,
we noticed that DeepVS behaved in a similar manner when the input from Dock6.6 was used.
In the experiments,
we varied one of the hyper-parameters and fixed all the others to the following values:
$d^{atm/amino/chrg/dist}=200$,
$cf=400$,
$\lambda=0.075$,
$k_{c}=6$
and $k_{p}=2$.

In Table~\ref{tbl:embedding_size},
we present the experimental results of varying the basic feature embedding sizes.
We can see that embedding sizes larger than 50 improved mainly the $ef$.
Embeddings larger than 200 did not improve the results.

\begin{table}[!htb]%\footnotesize
  \caption{DeepVS sensitivity to basic feature embedding sizes.}
  \label{tbl:embedding_size} 
	\begin{tabular}{lllll}
		\hline
         $d^{atm/amino/chrg/dist}$ & ef$_{max}$ & ef$_{2\%}$ & ef$_{20\%}$ & auc \\
		\hline
		 50	 & 15.4 & 6.5 & 3.0 & 0.803\\
		 100 & 15.6 & 6.8 & 3.1 & 0.799 \\
		 200 & 16.0 & 6.6 & 3.1 & 0.807 \\
    \hline
  \end{tabular}
\end{table}

The experimental results of varying the number of filters ($cf$) in the convolutional layer are presented in Table~\ref{tbl:conv_layer_size}.
Notice that the AUC improves by increasing the number of convolutional filters up to 400.
On the other hand,
using $cf=200$ results in the best $ef_{2\%}$.

\begin{table}[!htb] %\footnotesize
  \caption{DeepVS sensitivity to \# of filters in the convolutional layer.}
  \label{tbl:conv_layer_size} 
	\begin{tabular}{lllll}
		\hline
         $cf$ & ef$_{max}$ & ef$_{2\%}$ & ef$_{20\%}$ &auc \\
		\hline
		 100 & 14.7 & 6.4 & 3.0 & 0.797 \\
		 200 & 16.3 & 7.3 & 3.0 & 0.799 \\
		 400 & 16.0 & 6.6 & 3.1 & 0.807 \\
		 800 & 16.8 & 7.0 & 3.1 & 0.804 \\ 
    \hline
  \end{tabular}
\end{table}

In Table~\ref{tbl:learning_rate},
we present the experimental results of DeepVS trained with different learning rates.
We can see that larger learning rates work better for the DUD dataset.
A learning rate of 0.1 resulted in the best outcomes in terms of AUC and $ef$.

\begin{table}[!htb]%\footnotesize
  \caption{DeepVS sensitivity to the learning rate.}
  \label{tbl:learning_rate} 
	\begin{tabular}{lllll}
		\hline
         $\lambda$ & ef$_{max}$ & ef$_{2\%}$ & ef$_{20\%}$ & auc \\
		\hline
	 	  0.1   & 17.3 & 6.9 & 3.2 & 0.809\\
		  0.075 & 16.0 & 6.6 & 3.1 & 0.807\\
		  0.05  & 15.5 & 6.6 & 3.0 & 0.801\\
		  0.025	& 14.7 & 6.4 & 3.0 & 0.795\\ 
		  0.01  & 15.7 & 6.2 & 3.1 & 0.800\\ 
    \hline
  \end{tabular}
\end{table}

We also investigated the impact of using a different number of neighbors atoms from the compound ($k_{c}$) and the receptor ($k_{p}$).
In Table~\ref{tbl:num_neighbor},
we present some experimental results where we vary both $k_{c}$ and $k_{p}$.
For instance,
with $k_{c}=0$ and $k_{p}=5$,
it means that no information from the compound is used,
while information from the 5 closest atoms from the receptor is used.
In the first half of the Table,
we keep $k_{p}$ fixed to 5,
and vary $k_{c}$.
In the second half of the Table,
we keep $k_{c}$ fixed to 6,
and vary $k_{p}$.
As we can notice in the first half of Table~\ref{tbl:num_neighbor},
by increasing the number of neighbor atoms we use from the compound ($k_{c}$),
we significantly increase both $ef$ and AUC.
In the second half of the Table,
we can notice that using $k_{p} > 2$ degrades DeepVS-ADV performance.
As we conjecture in the next section,
this behavior seems to be related to the quality of the docking program output.

% \begin{table}[!htb]%\footnotesize
%   \caption{DeepVS sensitivity to \# of neighbor atoms selected from molecule/protein.}
%   \label{tbl:num_neighbor_1} 
% 	\begin{tabular}{lllll}
% 		\hline
%          $k$ & ef$_{max}$ & ef$_{2\%}$ & ef$_{20\%}$ & auc \\
% 		\hline
% 		  2 & 16.2 & 6.6 & 2.9 & 0.772 \\
% 		  3 & 16.4 & 6.1 & 2.8 & 0.776\\
% 		  4 & 15.3 & 6.0 & 3.0 & 0.784\\
% 		  5 & 14.0 & 5.9 & 3.0 & 0.790\\ 
%     \hline
%   \end{tabular}
% \end{table}

\begin{table}[!htb]%\footnotesize
  \caption{DeepVS sensitivity to \# of neighbor atoms selected from compound/protein.}
  \label{tbl:num_neighbor} 
	\begin{tabular}{cccccc}
		\hline
         $k_{c}$ & $k_{p}$ & ef$_{max}$ & ef$_{2\%}$ & ef$_{20\%}$ & auc \\
		\hline
		 0 & 5 &  6.99 & 2.18 & 1.54 & 0.574  \\
		 1 & 5 & 15.55 & 4.22 & 2.27 & 0.697  \\
		 2 & 5 & 15.44 & 5.62 & 2.59 & 0.743  \\
		 3 & 5 & 16.74 & 6.38 & 2.76 & 0.752  \\ 
		 4 & 5 & 17.38 & 6.25 & 2.91 & 0.782  \\
		 5 & 5 & 19.07 & 6.47 & 3.18 & 0.799  \\
		 6 & 5 & 17.89 & 6.79 & 3.04 & 0.799  \\
	     6 & 4 & 17.02 & 6.38 & 3.17 & 0.801  \\
	     6 & 3 & 16.44 & 6.82 & 2.99 & 0.793  \\
		 6 & 2 & 16.03 & 6.62 & 3.14 & 0.807  \\
		 6 & 1 & 16.03 & 6.99 & 3.13 & 0.806  \\
		 6 & 0 & 16.92 & 6.95 & 3.06 & 0.803  \\
    \hline
  \end{tabular}
\end{table}

In order to assess the robustness of DeepVS with regard to the initialization of the weight matrices (network parameters),
we performed 10 different runs of DeepVS-ADV using a different random seed in each run.
In Table~\ref{tbl:diff_seeds},
we present the experimental results of these 10 runs.
We can see in the Table that the standard deviation is very small for both $ef$ and AUC,
which demonstrates the robustness of DeepVS to different random seeds.

\begin{table}[!htb]%\footnotesize
  \caption{DeepVS sensitivity to different random seeds.}
  \label{tbl:diff_seeds} 
	\begin{tabular}{rrrrr}
		\hline
         Run & ef$_{max}$ & ef$_{2\%}$ & ef$_{20\%}$ & auc \\
		\hline
		 $average$ & 15.97 & 6.82 & 3.10 & 0.805 \\
         $stdv$    &  0.73 & 0.16 & 0.04 & 0.003 \\
		\hline
		
		  1 & 16.53 & 6.88 & 3.19 & 0.807 \\
		  2 & 16.32 & 6.63 & 3.06 & 0.799 \\
		  3 & 15.30 & 6.76 & 3.10 & 0.805 \\
		  4 & 14.92 & 6.89 & 3.12 & 0.807 \\
		  5 & 17.08 & 6.74 & 3.10 & 0.807 \\
		  6 & 16.82 & 6.84 & 3.05 & 0.804 \\
		  7 & 15.90 & 6.52 & 3.07 & 0.803 \\
		  8 & 15.59 & 7.00 & 3.06 & 0.805 \\
		  9 & 16.08 & 6.86 & 3.12 & 0.806 \\
		 10 & 15.18 & 7.04 & 3.11 & 0.803 \\
		 
    \hline
  \end{tabular}
\end{table}

\subsection{Docking Quality vs. DeepVS Performance}

In the experimental results reported in Table~\ref{tbl:num_neighbor},
we can notice that,
when creating the atom contexts,
using $k_{p} > 2$ (number of neighbor atoms coming from the protein) does not lead to improved AUC or $ef$.
In fact,
if we use only information from the compound ($k_{p} = 0$),
which is equivalent to perform ligand based virtual screening,
the AUC is already very good (0.803).
We hypothesize that this behavior is related to the quality of the docking output, 
which varies a lot across the 40 DUD proteins.
In an attempting to test this hypothesis,
we separately analyse the AUC of DeepVS for the DUD proteins for which the AUC of ADV is good  (Table~\ref{tbl:compare_best}) or poor (Table~\ref{tbl:compare_worse_auc}).

In Table~\ref{tbl:compare_best},
we present the AUC of DeepVS-ADV 
for proteins whose AUC of ADV is larger than 0.75.
We present results for three different values of $k_{p}$,
namely,
0, 2 and 5.
In the three experiments we use $k_{c}=6$.
We can notice that for proteins that have a good docking quality (and likely the protein-compound complexes have good structural information) the average AUC of DeepVS-ADV increases as we increase $k_{p}$.
This result suggests that if the structural information is good,
the neural network can benefit from it.

\begin{table}[!htb]%\scriptsize
  \caption{DeepVS-ADV results for proteins with good docking quality.}
  \label{tbl:compare_best} 
	\setlength{\tabcolsep}{4pt}
	\begin{tabular}{lccc}
		\hline
		 &\multicolumn{3}{c}{auc > 0.75} \\
         \cline{2-4} 
         & $k_{p}=0$ & $k_{p}=2$ &  $k_{p}=5$  \\
         \hline
        
		 $average$ & 0.83 & 0.86 & 0.87 \\

		 \hline
	
		 COX1           & 0.78 & 0.77 & 0.80   \\
		 COX2           & 0.89 & 0.91 & 0.91  \\
		 DHFR           & 0.96 & 0.94 & 0.96  \\
		 ER$_{agonist}$ & 0.89 & 0.88 & 0.89  \\
		 GART           & 0.78 & 0.77 & 0.77  \\
		 MR             & 0.80 & 0.82 & 0.80  \\
		 RXRa           & 0.64 & 0.85 & 0.90  \\		  
		 SAHH           & 0.93 & 0.95 & 0.95  \\		  
		  
  \hline
  \end{tabular}
\end{table} 

In Table~\ref{tbl:compare_worse_auc},
we present the AUC of DeepVS-ADV 
for proteins whose AUC of ADV is smaller than 0.5.
For these proteins,
which have a poor docking quality (and likely the protein-compound complexes have poor structural information) the average AUC of DeepVS-ADV decreases as we increase $k_{p}$.
This result suggests that if the structural information is poor,
the neural network works better without using it.

\begin{table}[!htb]%\scriptsize
  \caption{DeepVS-ADV results for proteins with poor docking quality.}
  \label{tbl:compare_worse_auc} 
	\setlength{\tabcolsep}{4pt}
	\begin{tabular}{lccc}
		\hline
		 &\multicolumn{3}{c}{auc < 0.50} \\
         \cline{2-4} 
         & $k_{p}=0$ &  $k_{p}=2$ & $k_{p}=5$  \\

         \hline
        
	 $average$ & 0.80 & 0.78 & 0.77 \\

		 \hline
		
		 ACE            & 0.71 & 0.72 & 0.66 \\
		 ADA            & 0.80 & 0.83 & 0.83 \\
		 AmpC           & 0.59 & 0.42 & 0.46 \\
		 COMT           & 0.92 & 0.92 & 0.89 \\
		 FGFrl          & 0.83 & 0.85 & 0.79 \\
		 HMGR           & 0.97 & 0.96 & 0.96 \\
		 NA             & 0.67 & 0.68 & 0.66 \\
		 PDGFrb         & 0.91 & 0.91 & 0.91 \\		  
		 PR             & 0.81 & 0.77 & 0.79 \\

  \hline
  \end{tabular}
\end{table}

\subsection{Comparison with State-of-the-art System}

In this section, we compare DeepVS results with the ones reported in previous work that also employed DUD.
First, we perform a detailed comparison between DeepVS and the Docking Data Feature Analysis (DDFA) system \cite{Arciniega2014},
which also applied neural networks on the output of docking programs to perform virtual screening.
Next,
we compare the average AUC of DeepVS with one of other systems that also report results for the 40 DUD receptors.

\begin{figure}[!htb]
    \includegraphics[scale=0.4]{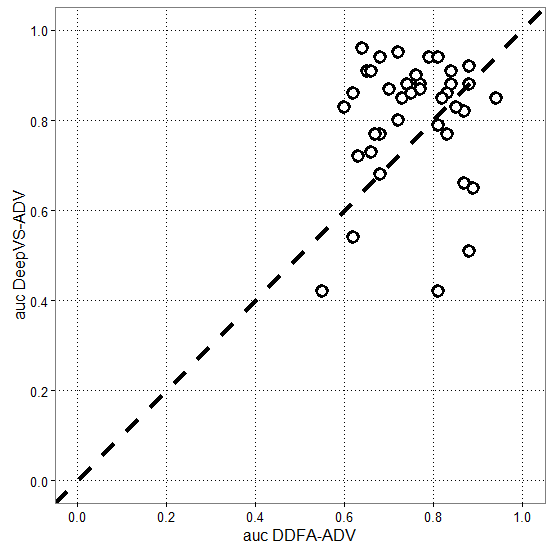}
	\caption{DeepVS-ADV vs DDFA-ADV.} 
  \label{fgr:deepvs_vs_ddfaadv}
\end{figure}

DDFA uses a set of human defined features that are derived from docking output data.
Examples of the features employed in DDFA are: compound efficiency, the best docking score of the compound poses and the weighted average of the best docking scores of the five most similar compounds in the docked library.
The features are given as input to a shallow neural network that classifies the input protein-compound complex as an active or inactive ligand.
DDFA uses data from the six best poses output by the docking program,
while in DeepVS we use data from the best pose only.
In Figure \ref{fgr:deepvs_vs_ddfaadv},
we compare the AUC of DeepVS-ADV vs DDFA-ADV,
which is the version of DDFA that uses the output of AutodockVina.
In this figure,
each circle represents one of the 40 DUD receptors.
DeepVS-ADV produces higher AUC than DDFA-ADV for 27 receptors,
which represents 67.5\% of the dataset.

\begin{figure}[!htb]
	\includegraphics[scale=0.4]{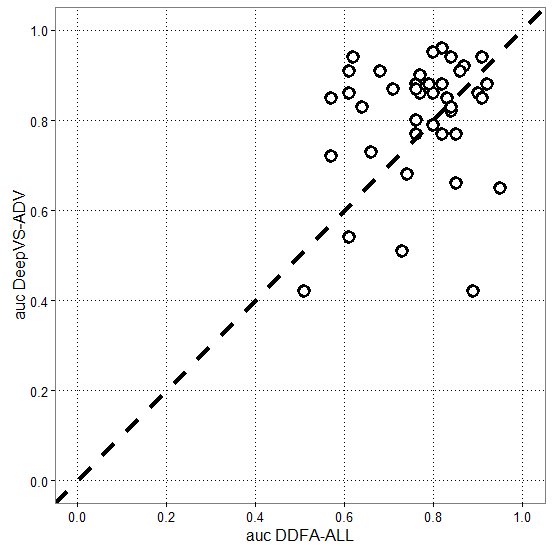}
	\caption{DeepVS-ADV vs DDFA-ALL.} 
  \label{fgr:deepvs_vs_ddfaall}
\end{figure}

DDFA-ALL is a more robust version of DDFA\cite{Arciniega2014} that uses simultaneously the output of three different docking programs: 
Autodock4.2 (AD4), 
AutodockVina1.1.2 (ADV)
and RosettaLigand3.4 (RL).
Therefore,
DDFA-ALL uses three times more input features than DDFA-ADV.
In Figure \ref{fgr:deepvs_vs_ddfaall},
we compare the AUC of DeepVS-ADV vs DDFA-ALL.
Despite using data from one docking program only,
DeepVS-ADV produces higher AUC than DDFA-ALL for 25 receptors,
which represents 62.5\% of the dataset.
This is a strong indication that DeepVS-ADV result for DUD dataset is very robust.

In Table \ref{tbl:state_of_the_art},
we compare the average AUC of DeepVS and the docking programs with the ones from other systems reported in the literature.
DeepVS-ADV produced the best AUC among all systems,
outperforming commercial docking softwares ICM and Glide SP.
NNScore1-ADV and NNScore2-ADV are also based on shallow neural networks that use human defined features and the output of AutodockVina.
It is worth to note that NNScore1-ADV and NNScore2-ADV \cite{Durrant2013} results are based in a different set of decoys that are simpler than the ones available in DUD.
Therefore,
these results are not 100\% comparable with other results on the table.
To the best of our knowledge,
the AUC of DeepVS-ADV is the best reported so far for virtual screening using the 40 receptors from DUD.

\begin{table}[!h]%\footnotesize
  \caption{Reported performance of different systems on DUD.}
  \label{tbl:state_of_the_art} 
  \begin{threeparttable}[b]
	\begin{tabular}{lllll}
		\hline
         System &&&auc \\
		\hline
		  \bf DeepVS-ADV       &&& \bf 0.81 \\
		  ICM \cite{Neves2012} \tnote{b} &&&0.79  \\
		  NNScore1-ADV \cite{Durrant2013} \tnote{a} &&&0.78\\
		  Glide SP \cite{Cross2009} \tnote{b} &&& 0.77\\
		  DDFA-ALL \cite{Arciniega2014}&&& 0.77   \\
		  DDFA-RL \cite{Arciniega2014} &&& 0.76  \\ 
		  NNScore2-ADV \cite{Durrant2013} \tnote{a} &&& 0.76\\
		  DDFA-ADV \cite{Arciniega2014} &&& 0.75\\
		  \bf DeepVS-Dock &&& \bf 0.74\\
		  DDFA-AD4 \cite{Arciniega2014} &&& 0.74\\
		  Glide HTVS \cite{Durrant2013} \tnote{a}&&& 0.73\\
		  Surflex \cite{Cross2009} \tnote{b} &&& 0.72 \\
		  Glide HTVS \cite{Cross2009} &&&  0.72 \\
		  ICM \cite{Neves2012} &&&  0.71 \\
		  RAW-ALL \cite{Arciniega2014} &&&  0.70 \\
		  Autodock Vina \cite{Durrant2013} \tnote{a} &&& 0.70\\
		  Surflex \cite{Cross2009} &&&  0.66 \\
		  RosettaLigand \cite{Arciniega2014} &&&  0.65 \\
		  Autodock Vina \cite{Arciniega2014} &&&  0.64 \\
		  ICM \cite{Cross2009} &&&  0.63 \\
		  \bf  Autodock Vina  &&&  \bf 0.62 \\
		  FlexX \cite{Cross2009} &&&  0.61 \\
		  Autodock4.2 \cite{Arciniega2014} &&&  0.60 \\
		  PhDOCK  \cite{Cross2009} &&&  0.59 \\
		  Dock4.0 \cite{Cross2009} &&&  0.55 \\
		  \bf Dock6.6 &&& \bf  0.48 \\
         \hline
  \end{tabular}
 \begin{tablenotes}
  \item[a]{{\footnotesize Used a different dataset of decoys.}}
  \item[b]{{\footnotesize Tuned by expert knowledge.}}
\end{tablenotes}	
  \end{threeparttable}
\end{table}

\section{Conclusions}
In this work, we introduce DeepVS,
a deep learning approach to improve the performance of docking-based virtual screening.
Using DeepVS on top of the docking output of AutodockVina we were capable of producing the best AUC reported so far for virtual screening on the DUD dataset.
This result, 
together with the fact that (1) DeepVS does not require human defined features, 
and (2) it achieves good results using the output of a single docking program,
makes DeepVS an attractive approach for virtual screening.
Moreover,
different from other methods that use shallow neural networks with few parameters,
DeepVS has a larger potential for performance improvement if more data is added to the training set.
Deep Learning systems are usually trained with large amounts of data.
Although the number of protein-compound complexes in DUD is relatively large (more than 100k),
the number of different proteins is still very small (only 40).

Additionally,
this work also brings some very innovative ideas on how to model the protein-compound complex raw data to be used in a deep neural network.
We introduce the idea of atom and amino acid embeddings, 
which can also be used in other deep learning applications for bioinformatics.
Moreover,
our idea of modeling the compound as a set of atom contexts that is further processed by a convolution layer proved to be an effective approach to learning representations of protein-compound complexes.

\begin{acknowledgement}
J.C.P. is funded through a Ph.D. scholarship from the Oswaldo Cruz Foundation. E.R.C's research is supported by following grants: Faperj/E-26/111.401/2013 and CNPq. Papes VI./407741/2012-7. 

\end{acknowledgement}

%\begin{suppinfo}

%This will usually read something like: ``Experimental procedures and
%characterization data for all new compounds. The class will
%automatically add a sentence pointing to the information on-line:

%\end{suppinfo}

\bibliography{deep_bio_jcim}

\end{document}